\documentclass{llncs}
\usepackage{graphicx,fancyvrb}

\usepackage{latexsym}
\usepackage{amssymb}
\usepackage{amsmath}
\usepackage{amsfonts}
\usepackage{proofs}
\usepackage[hyphens]{url}

\usepackage{hyperref}

\usepackage[table,xcdraw]{xcolor}
\usepackage{multirow}
\usepackage{listings}
\usepackage{xcolor}

\usepackage{tikz}
\usepackage{subfigure}
\usetikzlibrary{mindmap,trees,backgrounds}

\usepackage{rotating}
\usepackage{algorithmic}
\usepackage{algorithm}

\usetikzlibrary{shapes,arrows,decorations.pathmorphing,backgrounds,fit,positioning,shapes.symbols,chains,shadings}
\newcommand{\myunit}{0.1 cm}
\tikzset{node style sp/.style={draw,circle,minimum size=\myunit}}

\begin{document}

\title{An Argumentation-Based Reasoner to Assist Digital Investigation and Attribution of Cyber-Attacks}

\author{Erisa Karafili
\and
Linna Wang
\and Emil C. Lupu}
\institute{Department of Computing, Imperial College London\\
180 Queen’s Gate, SW7 2AZ, London, UK\\ \email{\{e.karafili, linna.wang15, e.c.lupu\}@imperial.ac.uk}}

\maketitle

\begin{abstract}
We expect an increase in the frequency and severity of cyber-attacks that comes along with the need for efficient security countermeasures. The process of attributing a cyber-attack helps to construct efficient and targeted mitigating and preventive security measures. In this work, we propose an argumentation-based reasoner (\emph{ABR}) as a proof-of-concept tool that can help a forensics analyst during the analysis of forensic evidence and the attribution process. Given the evidence collected from a cyber-attack, our reasoner can assist the analyst during the investigation process, by helping him/her to analyze the evidence and identify who performed the attack. Furthermore, it suggests to the analyst where to focus further analyses by giving hints of the missing evidence or new investigation paths to follow. \emph{ABR} is the first automatic reasoner that can combine both technical and social evidence in the analysis of a cyber-attack, and that can also cope with incomplete and conflicting information. To illustrate how \emph{ABR} can assist in the analysis and attribution of cyber-attacks we have used examples of cyber-attacks and their analyses as reported in publicly available reports and online literature. We do not mean to either agree or disagree with the analyses presented therein or reach attribution conclusions. 
\end{abstract}

%


\section{Introduction}
The increase in cyber-attacks we are currently 
facing~\cite{WebNewman2017} is expected to continue, especially given the exponential increase in the usage of IoT and smart devices, which drastically increases the attack surface of systems.  The increasing dependency users have on these connected devices raises the users' exposure to cyber-attacks. The growth in frequency and severity of cyber-attacks comes along with the increased economic costs associated to the damages caused by such cyber-attacks~\cite{csoonline}.
Existing protective and mitigating measures are not sufficient to cope with the sophistication of current attacks. This brings the need to enforce efficient preventive and mitigating measures that are attacker-oriented, i.e., countermeasures 
that are specific to the attacker or group of attackers performing the attack. Furthermore, discovering who performed an attack and bringing the perpetrators to justice, can act as a deterrent for future cyber-attacks.

Attacker-oriented countermeasures require to discover the perpetrator 
of the attack or the entity related to it.  \emph{Attribution} is the process of assigning an action of a cyber-attack to a particular entity/attacker/group of attackers. Currently, the attribution of cyber-attacks is mainly a manual process,
performed by the forensic analyst, and is strictly related to the knowledge of the analyst, thus, is easily 
human biased and error-prone. Attributing cyber-attacks is not trivial, as attackers often use deceptive and anti-forensics 
techniques~\cite{goutam15}, and the analysts need to analyze an enormous amount
 of data, filter~\cite{KarafiliCV18,RasgaSKV19} and classify them.
The increasing use of IoT devices aggravates the work of the analysts
 and makes the attribution process more expensive, as the analysts 
 might need to physically access the devices to retrieve their data.

Digital forensics helps during the attribution process,
as it collects and analyzes the evidence left by the attack,
but it is not able to deal with conflicting or incomplete information.
It only works with technical evidence, and fails to consider other aspects such as geopolitical 
 situations and social-cultural contexts that provide useful leads during an investigation.
Digital forensics tools mainly focus on collecting the evidence,
which is then given to the analyst for analysis. This 
makes the process often extremely human-intensive, requiring many 
skilled analysts to work for weeks or even months~\cite{nassif13,openlv}.
 The problem is aggravated by the large proportion of unstructured 
 data, which makes the automated analysis challenging.

In this work, we propose an automatic \emph{reasoner} (\emph{ABR}), based
on argumentation and abductive reasoning that helps the forensic analyst during 
the evidence analysis and attribution process. Given the pieces of cyber forensic and social evidence
 of a cyber-attack, the proposed reasoner analyzes them and derives new information that is provided to the
analyst. In particular, \emph{ABR} can answer queries, such as, who is a possible perpetrator 
of an attack, who has the motives to perform it, what are the capabilities needed to perform an attack or what are the similarities with past attacks. Furthermore, \emph{ABR} can suggest to the analyst other paths of investigation, by giving hints on what other pieces of evidence can be collected to arrive at a conclusion thus, enabling a prioritized evidence collection. 
Our reasoner is based on our preliminary work~\cite{KarafiliWKL18,SocialGood}  where we briefly presented the main intuition behind \emph{ABR}. To the best of our knowledge, this is the first automatic reasoner that helps with the analysis of cyber-attacks using both technical and social evidence, and that is able to reason with conflicting and incomplete knowledge.

The reasoner uses a set of \emph{reasoning rules}, preferences between them,
and \emph{background knowledge}. 
The rules of the reasoner are constructed from the input provided by the expert user and from the analyses of past attacks. To illustrate \emph{ABR}, we have used in this paper rules extrapolated from the analysis of well-known cyber-attacks as published in the public literature (e.g., APT1~\cite{Mandiant2013ExposingUnits}, Wannacry~\cite{Wannacry}).
In particular, the rules were generalised so that they can be applied across different attack scenarios. The background knowledge incorporates typical common knowledge that analysts may use during the analysis process. In this work, we have used some of the knowledge extracted from the examples, as well as other public reports such as \cite{GCI2017,WebBreene2016WhoSuperpowers,WebBrilliantMaps2017WhoEnemies}.  
Our reasoner can assist in the attribution of an attack by using both technical evidence and social considerations that are represented thanks to the use of a social model~\cite{rid15}.
 
\emph{ABR} is able to work with incomplete and conflicting evidence.
We decided to base \emph{ABR} on an \emph{argumentation} framework, in particular, a preference-based argumentation
framework~\cite{KMD94}, which permits to reason with \emph{conflicting} pieces of evidence
by introducing preferences between the applied rules.
We use preference-based argumentation as it is similar to the decision-making 
 process followed by digital forensics investigators. 
\emph{ABR} is constructed using the 
Gorgias~\cite{gorgias1} tool, which uses 
abductive reasoning~\cite{abductive} combined with preference-based 
argumentation. The use of \emph{abduction} allows us to reach conclusions even with \emph{incomplete} information, 
as the missing information is abduced (hypothesized) and then suggested to the analyst as hints of possible further evidence to be collected.

\emph{ABR} is a proof-of-concept tool that aims to assist the analyst during the 
analysis process. Therefore, together with the answer to a query 
it also provides the explanation of the reasoning process, applied rules 
and the information used to reach that conclusion. 
Furthermore, \emph{ABR} gives hints to the analyst for missing evidence,
that, if provided, allows to pursue other investigation paths. 
\emph{ABR} is flexible and adaptable to user requests and 
changes. The use of \emph{ABR} helps to promote best practices and to share lessons learned from past experience as rules and background knowledge can be constructed with expert input and then shared and re-used across investigations. 

In Section~\ref{relatedwork} we present the relevant related work. 
We introduce our argumentation-based reasoner (\emph{ABR}) in Section~\ref{ABR}.
In Section~\ref{sec:rules} and \ref{sec:bg} 
we present \emph{ABR}'s main components,
correspondingly its reasoning rules and its background knowledge. 
We give an overall evaluation and discussion in Section~\ref{discuss}.
In Section~\ref{concl} we conclude and present some interesting future research directions.

\section{Related Work}\label{relatedwork}
Attribution of a cyber-attack is the process of 
``determining the identity or location of an attacker or attackers intermediary''~\cite{wheeler03}. 
Tracing  the origin of a cyber-attack is difficult as attackers can easily forge or obscure information sources, and use anti-forensics tools, to avoid being detected and identified~\cite{goutam15}.
Digital forensics plays a significant role in attribution by collecting, examining, analyzing and reporting 
 the evidence~\cite{Kent2006GuideResponse}.
Other techniques created for protecting the systems are also used to collect 
forensic data, e.g., traceback techniques~\cite{wheeler03},
honeypots~\cite{anagnostakis2005}, or other deception 
techniques~\cite{Almeshekah2014,Almeshekah2016,Val2017}. 

Digital forensics comes with its own challenges~\cite{Beebe}, that can mainly be categorised into:
\emph{complexity problems} as the collected data are in the lowest raw format and 
require high resources to analyze them and 
\emph{quantity problems} as the enormous amount of collected data is too large to be analyzed 
manually~\cite{carrier03}. 
Forensics techniques identify and collect the evidence that is later managed and analyzed 
by the forensic analyst.
Since often the data are collected from different sources and
the attackers can plant false evidence to lead the investigator off his/her trail, 
the latter is likely to be in a situation with multiple pieces of conflicting evidence.
Digital forensics techniques can deal with conflicting information
during the evidence collection phase~\cite{aziz,fontani},
but lack the ability to work with 
conflicting pieces of evidence during the analysis and attribution process.  
These techniques can collect pieces of evidence~\cite{schatz},
but have difficulties reasoning with 
 incomplete information, and reaching conclusions
 without having all the needed pieces of evidence.  
Digital forensics only uses technical evidence~\cite{openlv} and fails to consider other factors such as 
geopolitical situations and social-cultural contexts, which could provide useful leads during the 
investigations.

A theoretical social science model is proposed in~\cite{rid15}, called the \emph{Q-Model} that describes how
the analysts combine technical and social evidence during the attribution process.
In this model, attribution is described as an \emph{incremental process} passing from one level of attribution 
to the other. The \emph{Q-Model} represents how the forensic investigators 
perform the attribution process and particular attention is placed on the social evidence,
where contextual knowledge such as ongoing conflicts between countries or rivalry between corporations are very useful in detecting 
motives of potential culprits.

We decided to use argumentation for our reasoner, as argumentation helps
during the analysis and attribution process
because it 
is transparent and encourages the evaluation of the arguments, by assessing the relative importance of various 
factors when making decisions~\cite{ouerdane10}. 
Argumentation captures the fact that the final decision might change if more information is available (i.e., 
\textit{non-monotonic reasoning}~\cite{nonmono}),
where more information may reveal new arguments that 
are in conflict with the original ones and are stronger than them. 
Non-monotonic reasoning has previously been proposed to tackle the attribution challenge. For example, in~\cite{nunes16,shakarian15} the authors propose the DeLP3E framework to attribute operations of cyber-attacks. 
This theoretical framework is based on the extension of Defeasible Logic Programming with probabilistic uncertainty. 
The DeLP3E framework does not deal with incomplete evidence, and thus, cannot make assumptions to reach a conclusion and cannot suggest new paths of investigation or new evidence to be collected. 
It also lacks in general technical and social common knowledge, e.g., ongoing conflicts/rivalries 
between countries/corporations, information about past attacks, cyber-security capabilities 
of entities, which can be very useful in detecting motives, capabilities and potential culprits. 
DeLP3E uses as a measure for its conclusion the probabilities of an event being true. 
However, this requires the user to provide the probability of being true for each of the given pieces of evidence.
It also does not distinguish the different levels of reasoning that can be applied to reach certain conclusions.

Despite the advances in using digital forensics or defeasible reasoning in attribution, some shortcomings still remain to be addressed. The most important one is that none of the current works considers the social aspects of attribution.
The current state of the art does not deal with incomplete evidence, which is an important aspect of forensic investigations, as usually not all evidence can be collected due to time/resource constraints, and anti-forensics tools used by attackers can hide some of the evidence. 
We believe, our reasoner is the first attempt to use a social model to categorize evidence and rules in an 
argumentation-based framework, which leads to a more accurate and explainable attribution
that helps the investigator during the analysis process also in case of conflicting and incomplete evidence.

\section{Argumentation-Based Reasoner for Attribution}\label{ABR}

\begin{figure}[htb]
\centering
\includegraphics[width=0.65\textwidth]{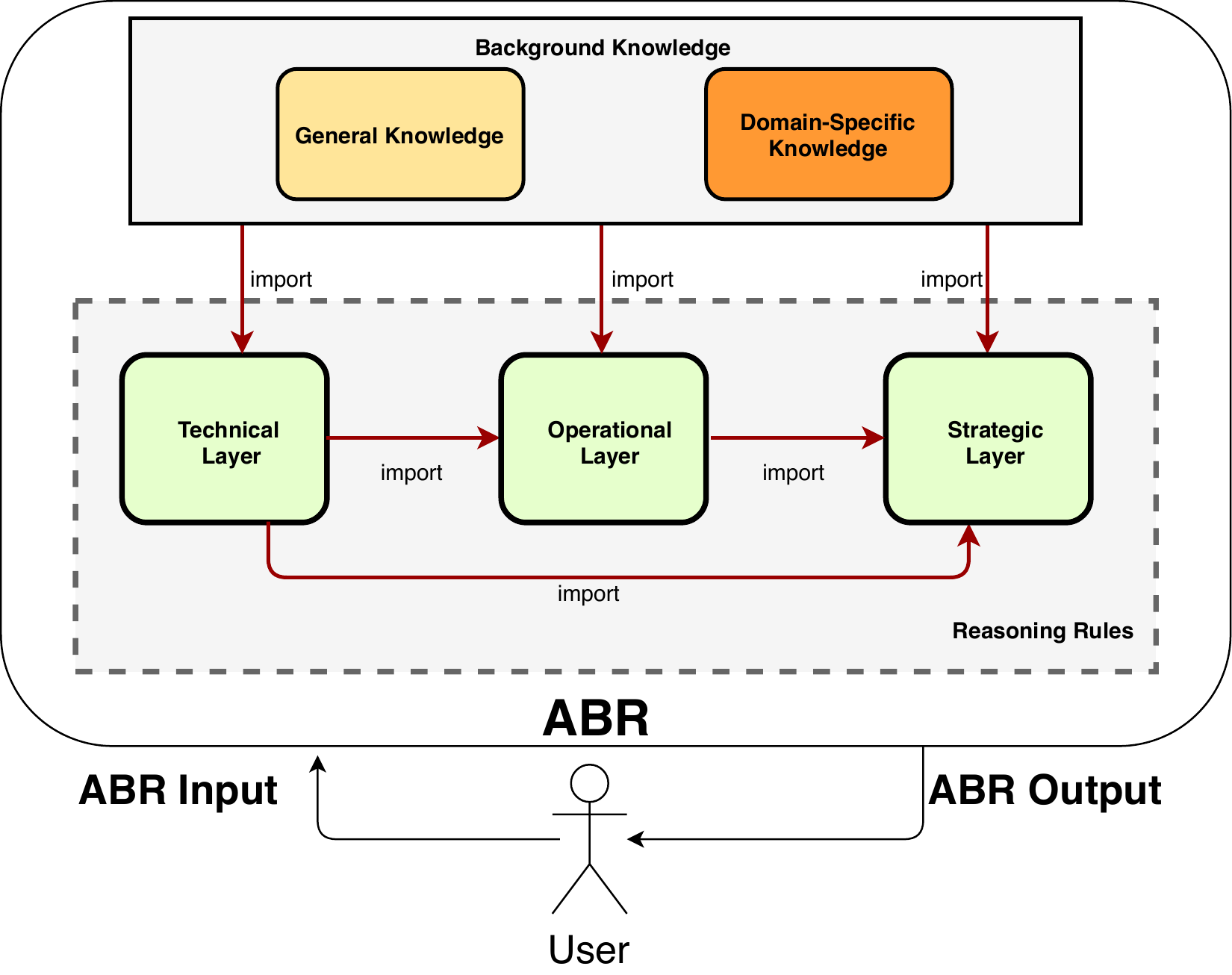}
\caption{\emph{ABR} Overview\label{fig:abr}}
\end{figure}

Let us now introduce our argumentation-based reasoner (\emph{ABR})
that is based on a preference-based argumentation framework. 
\emph{ABR} is composed of two main components the
 \emph{reasoning rules}, and the \emph{background knowledge},
 see Figure~\ref{fig:abr}.
Given the evidence presented in input, \emph{ABR} analyzes it
and attempts to answer queries about the possible perpetrators of the attack, or
provides suggestions for further pieces of evidence needed to reach a conclusion or perform a more precise or a different analysis.
The reasoning rules used by ABR were 
 extracted\footnote{Currently the extraction of the rules is done manually by analyzing various
reports and articles about the analysis and attribution of past cyber-attacks.} from public reports about past cyber-attacks and formalized in the argumentation framework\footnote{The rules extracted have not been evaluated for correctness and might not be complete i.e., they might not capture the complexity of the situations encountered.}. 
In actual use, rules could be
 specified by expert users or extracted automatically from different analyses and then reviewed by expert analysts. 
Rules are divided into three layers: technical,
operational and strategic layer, following the social model structure proposed in~\cite{rid15}. 
In this paper, we have used background knowledge based on the information extracted from online analyses of past cyber-attacks and 
relevant information for these attacks. 
\emph{ABR} takes as input from the user the pieces of evidence
(technical and social evidence) relevant to the current investigation and then 
analyzes them by using the reasoning rules and the background knowledge.
It gives as result to the user answers to the user's queries, e.g., 
if a given entity is a possible culprit of the attack,
together with an explanation on how the conclusion was reached, hints
about what other pieces of evidence the user can provide to perform a more precise or a new analysis. 

\subsection{Argumentation Framework for Attribution}\label{argum}
\noindent
We base our reasoner on
a preference-based 
argumentation framework~\cite{KMD94,gorgias1},
 as it permits the user to take decisions while working with conflicting evidence,
and it naturally encodes the different reasoning layers with its preference relations between rules. 
The used framework best simulates the analysis and 
attribution process made by an investigator, who needs to use different reasoning rules
that work with technical and social aspects of the attack, 
 have exceptions, and can derive conflicting conclusions.
 
Our framework allows the investigator to work with conflicting evidence and
reasoning rules that derive conflicting conclusions, 
by introducing preferences between them. 
The introduced preferences can be considered  as exceptions to other rules, or
preferences that are context dependent.  
The use of argumentation permits to provide an explanation of the given results.
Let us briefly introduce the used framework.

An \emph{argumentation theory} is a pair $(\mathcal{T}, \mathcal{P})$
of argument rules $\mathcal{T}$ and preference rules $\mathcal{P}$.
The \emph{argument rules} $\mathcal{T}$ are a set of labeled formulas of the form:
$$
 rule_i: L \leftarrow L_1, \ldots, L_n$$
where $L, L_1, \ldots, L_n$ are positive or negative ground  literals, and $rule_i$ is
the label denoting the rule name.
In the above argument rule, $L$ denotes the \emph{conclusion} of the argument
rule 
and $L_1, \ldots, L_n$ denote its \emph{premises}. 
The premise of an argument rule is the set of conditions required for the conclusion to be true. 
In our framework, the argument rules are the \emph{reasoning rules} used by \emph{ABR}. 
Let us show below
a reasoning rule that is part of \emph{ABR}: 
\begin{displaymath}
str_1:   isCulprit(C, Att) \leftarrow ClaimResp(C, Att)
\end{displaymath}
where the \emph{rule name} is the label of the rule,
in this case
$str_1$; 
the \emph{head} is the second argument and represents the conclusion of the rule, in this case $isCulprit(C, Att)$;
the \emph{body} predicates are the literals following the head, and represent the premises
of the rule, in this case $ClaimResp(C, Att)$.

The \emph{preference rules} $\mathcal{P}$ are a set of labelled formulas of the form: 
$$
 p_i: rule_1 > rule_2$$
where 
 $p_i$ is the label denoting the rule name,
the head of the rule is $rule_1 > rule_2$, 
and $rule_1$, $rule_2$ are labels of rules defined in $\mathcal{T}$,
and $>$ refers to an irreflexive, transitive and antisymmetric higher priority relation
between rules. 
The above rule means that $rule_1$ has \emph{higher priority} than $rule_2$,
 or better $rule_1$ is preferred over $rule_2$. 
The preference rules, also called \emph{priority rules} are true always or in certain conditions or contexts.
We show below a priority rule, $p_1$, denoting that rule $str_2$ is preferred over rule $str_1$. 
$$
 p_1: str_2 > str_1$$

We have priority rules between rules that are in conflict with each other or better that derive conflicting conclusions. 
\emph{Preference-based argumentation} allows the investigator to handle non-monotonic reasoning~\cite{nonmono} in attribution, where the introduction of new evidence might change the result of the attribution (due to conflicting arguments) and the investigator's confidence in the results. 
Argumentation is particularly useful as it permits to represent the 
reasoning rules in an intuitive and simple way. 

Let us introduce the following rule that is part of \emph{ABR}:
$$str_2: \neg isCulprit(X, Att) \leftarrow \neg hasCap(X, Att).$$
Rule $str_2$ describes that entity $X$ is not the possible culprit for the attack $Att$, because it does not have the capabilities 
for performing it. Rule $str_1$ and $str_2$ are in conflict with each other because when both preconditions are met, they derive conflicting conclusions. Given the above preference rule $p_1$, 
rule $str_2$ is preferred over rule $str_1$. Thus, in case both preconditions for  $str_1$ and $str_2$ are given,
we take into consideration only the conclusion from $str_2$ , $\neg isCulprit(C, Att)$.

The inputs of \emph{ABR} are pieces of evidence 
that are used together with the background knowledge by the reasoning rules to derive new information. 
The reasoning rules and the preferences used in this paper were extracted from 
real cyber-attacks analyses and attribution taken from online public reports,
 such as~\cite{Mandiant2013ExposingUnits,Wannacry}.

\emph{ABR} is the first tool that is able to work with incomplete evidence.
It provides hints of missing evidence or new investigation paths to the user, thanks to the use of abductive reasoning~\cite{abductive}.  
The use of abducible predicates
 permits to fill  the \emph{knowledge gaps} in the reasoning,
by allowing \emph{ABR} to perform the analysis and to reach a conclusion even when
 there are insufficient pieces of evidence.
 This feature is extremely important to the investigator who is provided with new possible conclusions and new evidence to be collected. 
 To construct \emph{ABR} we use the Gorgias~\cite{gorgias1} tool,
which is a preference-based argumentation reasoning tool that uses abduction.

Let us now introduce the following rule from \emph{ABR}: 
\begin{displaymath}
\begin{array}{ll}
op_1: hasMotive(X,Att) \leftarrow & \ target(T,Att),\ industry(T), \\ 
& hasEconMot(X,T),  \\ & contextOfAtt(econ,Att), \\ &
 specificTarget(Att).
\end{array}
\end{displaymath}
which states that $X$ has the motives to perform attack $Att$, when it has economical motives against the target $T$ of $Att$, where $T$  is an industrial company, the context of $Att$ was economical ($econ$), and $Att$ had a specific target. 
\emph{ABR} treats $specificTarget$ as an \emph{abducible} predicate. 
For every abducible predicate we have the rules that derive the predicate or its negation.
For the $specificTarget$ abducible we can prove that it is not true by using the following rule.
\begin{displaymath}
\begin{array}{ll}
op_2: \neg specificTarget(Att) \leftarrow & target(T_1,Att), \\
&  target(T_2,Att), T_1 \neq T_2. 
\end{array}
\end{displaymath}
In case, we are not able to derive $\neg specificTarget(Att)$, then we can abduce (hypothesize) that $specificTarget(Att)$ is true, and we can use this result
to derive $hasMotive(X, Att)$, in case we have the rest of the preconditions.

\subsection{Technical and Social Attribution}
The main goal of the \emph{ABR} reasoner is to assist the forensic analyst during the 
evidence analysis. 
Given the pieces of evidence of an attack, the reasoner
analyzes the evidence and derives new information, if possible attributes this attack to one or different possible entities, or 
provides suggestions on other pieces of evidence that the user can provide to better analyze and attribute the attack. 
To perform the attribution process, \emph{ABR} also needs to work with non-technical  evidence, usually called \emph{social evidence}. 
To deal with these aspects, we have used a social model for attribution, called the \emph{Q-Model}~\cite{rid15}. This model represents how the investigators perform the attribution process of cyber-attacks. Following the Q-Model, we categorize the evidence and the reasoning rules into three layers:  \textit{technical}, \textit{operational} and \textit{strategic}. 
 The combination of information in these layers permits the attribution of a cyber-attack, as it aims to emulate the investigator's attribution process. Depending on the layer a rule/evidence is part of, we call it a technical, operational, or strategic rule/evidence and denote its name starting correspondingly with
$t$, $op$, or $str$.

The \emph{technical layer} is composed of rules that deal with pieces of evidence obtained from digital forensics processes,
related to technical evidence of the attack, and how it was carried out, e.g., the IP address from which the attack was originated, 
time of the attack, logs, type of attack, code used. 
Let us
 give below an example of a technical layer reasoning rule that is part of \emph{ABR}:
$$t_1: reqHighRes(Att) \leftarrow usesZeroDay(Att).
$$
Rule $t_1$ denotes that if the attack $Att$ uses zero-day vulnerabilities, $usesZeroDay(Att)$, 
then this attack requires a lot of resources, $reqHighRes(Att)$.

The \emph{operational layer} is composed of rules that deal with
 non-technical pieces of evidence that relate to the social aspects where the attack took place, 
 e.g., the motives of the attack, the needed capabilities to perform it, 
 the political or economical context where it took place. 
Let us give below an operational layer reasoning rule that is part of \emph{ABR}:
\begin{displaymath}
\begin{array}{ll}
op_3: hasCap(X, Att) \leftarrow & reqHighRes(Att),\\ & hasResources(X).
\end{array}
\end{displaymath}
Rule $op_3$ denotes that if
 $Att$ requires a large amount of resources, and an entity $X$ has (large amounts of) resources,
$hasResources(X)$, 
then $X$ has the capability to carry out the attack, $hasCap(X, Att)$.

The \emph{strategic layer} is composed of rules that deal with who
 performed the attack, or who is obtaining advantage from it.
Let us give below a strategic layer reasoning rule that is part of \emph{ABR}:
 \begin{displaymath}
\begin{array}{ll}
str_3: isCulprit(X, Att) \leftarrow & hasMotive(X, Att),\\
& hasCap(X, Att).
\end{array}
\end{displaymath}
Rule $str_3$ denotes that if $X$ has both the capability, $hasCap(X, Att)$, and the motive, $hasMotive(X, Att)$, to carry out the attack $Att$,
then $X$ is a possible culprit of the attack, $isCulprit(X, Att)$.

As shown in Figure~\ref{fig:abr}, the operational rules use information derived from the technical layer, and the strategic rules use information derived
 from the technical and operational layers. All three layers use the evidence given by the user and the
 background knowledge. 
This categorization of the evidence and rules in three layers, following from
the Q-Model, aims to emulate the forensic investigator's analysis 
during the attribution process, where s/he moves from the technical layer,
to the operational, and finally to the strategic one, by using the conclusions from the previous layers. 
Furthermore, this categorization 
improves \emph{ABR}'s usability, given the investigator's familiarity with these three layers.

\section{\emph{ABR}'s Reasoning Rules}\label{sec:rules}
To illustrate the use of ABR we have extracted around 200 reasoning rules from the analyses of different cyber-attacks reported in the public literature (e.g., APT1~\cite{Mandiant2013ExposingUnits} and 
Wannacry~\cite{Wannacry}).
 These rules have then been translated into generic argumentation rules to be used within the framework. 
 These reasoning rules are considered as one of the main components of \emph{ABR} as they permit to perform the reasoning behind the analysis and attribution of cyber-attacks. 
 We briefly present some of these rules in this section. 

As described in the previous section, the  reasoning rules, also called simply rules, 
are divided into three layers: technical, operational and strategic.
Let us give an overview of some of the strategic rules
of the reasoner and show how the rules of the different 
layers are related to each other.
The following rules describe some of the circumstances in which we can derive that an entity $X$ is a possible culprit 
($isCulprit(X, Att)$) or not ($\neg isCulprit(X, Att)$) of an attack $Att$.

\begin{displaymath}
\begin{array}{ll}
str_3: \ isCulprit(X, Att) \leftarrow & hasMotive(X, Att),\\ & hasCap(X, Att).\\
str_4: \ isCulprit(X, Att) \leftarrow & malwareUsed(M1, Att),\\ &
 similar(M1, M2),\\
& notBlackMarket(M1), \\ &
notBlackMarket(M2),\\
& malwareLinked(M2, X). \\
str_5: \neg isCulprit(X, Att) \leftarrow &\neg attackOrig(X, Att).\\
str_6: \neg isCulprit(X, Att) \leftarrow &target(X, Att).
\end{array}
\end{displaymath}
Let us use the strategic rule $str_3$, to show the relations of the reasoning
 rules
between the different layers.
Rule $str_3$ uses the predicates $hasMotive(X, Att)$ and $hasCap(X, Att)$; 
where the first is a derived predicate of the operational layer, indicating that $X$
has motives to perform the attack $Att$ and $hasCap(X, Att)$ is a derived predicate of the technical and operational
layer, indicating that entity $X$ has the capabilities to perform $Att$. 

The $hasMotive$ predicate can be derived using the rule introduced in Section~\ref{argum}, represented as below:
\begin{displaymath}
\begin{array}{ll}
op_1: hasMotive(X,Att) \leftarrow &target(T,Att),\ industry(T), \\ & hasEconMot(X,T), \\&contextOfAtt(econ,Att), \\ &
 specificTarget(Att).
\end{array}
\end{displaymath}
The above rule says that an entity $X$ has the motives to perform $Att$,
when $X$ has economical motives to attack a particular entity $T$, which
is an industry, and the attack was designed to target entity $T$, 
and the context of $Att$ was economical. 
The predicates used in $op_1$ are:
$target(T, Att)$ is an evidence, stating that $T$ is the target of $Att$;
$industry(T)$ is a background fact, stating that $T$ is an industry;
$hasEconMot(X, T)$ is an evidence, stating that entity $X$
 benefits economically from attacking industry $T$, (for example, if $countryC$ has identified $industryY$ as a strategic 
 industry,  we say that $hasEconMot(countryC, industryY)$ is true);
$specificTarget(Att)$ is an evidence that is 
true when $Att$ was constructed to attack a particular target;
$contextOfAtt(Y, Att)$ is an evidence stating that: if the target of an attack was a ``normal''\footnote{``Normal'' industries are companies that are not closely related  to a country's national interests.
A ``political'' industry is a company that is closely related  to a country's national interests, e.g., the defence or energy sector.
} industry, 
then the context was economical 
($econ$), 
if the target was a ``political'' industry, then the context was political ($pol$). 

We introduce below one of the rules that derives the $hasCap$ predicate\footnote{Numerous factors can be used to determine the capability. However, for the sake
of space, we introduce only one of the possible rules that can derive the capability ($hasCap$) predicate.}. 
\begin{displaymath}
\begin{array}{ll}
op_3: hasCap(X, Att) \leftarrow & reqHighRes(Att), \\ & hasResources(X).
\end{array}
\end{displaymath}
Rule $op_3$ states that $X$ has the capability to perform $Att$, when $Att$ 
requires high resources and $X$ has the needed resources.
Predicate $reqHighRes(Att)$
can be derived from the following technical rules:
\begin{displaymath}
\begin{array}{ll}
t_2: reqHighRes(Att) \leftarrow &target(T, Att), \ highSecurity(T).\\
t_3: reqHighRes(Att) \leftarrow   &highVolAtt(Att),\ longDurAtt(Att).\\
t_4: reqHighRes(Att) \leftarrow  & highLevelSkill(Att).\\
\end{array}
\end{displaymath}
where
$highSecurity(T)$ means that entity $T$ has high security measures in place;
$highVolAtt(Att)$ means that $Att$ has a high volume;
$longDurAtt(Att)$ means that $Att$ was performed over a long duration (few months or even years),
and $highLevelSkill$ means that $Att$ is a complex attack and requires high level skills to be performed.
Rule $t_2$ states that $Att$ requires high resources if its target has put in place high
security measures, rule $t_3$ states that $Att$ requires high resources if the attack has a high 
volume and a long duration, and rule $t_4$ states that $Att$ requires high resources if it requires advanced skills.

\emph{ABR}'s rules are used to analyze the evidence and to derive new conclusions, in order to offer new insights to the analyst, as shown in the example below. 

\begin{example}\label{usbank}
Let us consider the example of the US bank hack~\cite{Goldman2012}, that
occurred in 2012. During this attack, US banks faced
denial of service (DoS) attacks,
 causing websites of many banks to suffer slowdowns 
 and even be unreachable for many customers.
The banks' web hosting services were infected
 by a sophisticated malware called \emph{Itsoknoproblembro}, (denoted as $itsOKnp$).
  Earlier that year, US government placed economic sanctions against Iran.
 Some of the pieces of evidence provided to \emph{ABR} for this attack ($usBHack$) are as below:
  \begin{displaymath}
 \begin{array}{l}
target(us\_banks, usBHack). \\ 
 targetCountry(usa , usBHack). \\ 
 attackPeriod(usBHack , [2012, 9]). \\
highLevelSkill(usBHack).\\
malwareUsed(itsOKnp, usBHack). \\
imposedSanc(usa, iran, [2012, 2]).
\end{array}
\end{displaymath}
By using rule $t_4$ and the evidence that this attack required a high level of skill, \emph{ABR} derives that this attack
requires high resources, $reqHighRes(usBHack)$.

Another rule capturing that an entity might have a political motive if it has been the target of sanctions can be written as follows:
\begin{displaymath}
\begin{array}{l}
op_4: hasPolMotive(C,T,Date) \leftarrow
imposedSanc(T,C,Date).
\end{array}
\end{displaymath}
Rule $op_4$ would then derive\footnote{\emph{ABR}'s derived evidence is derived from the application of the rules to the input evidence and thus depends on both rules and evidence being correct.} that Iran might have political motives against US 
because of the sanctions imposed by US against Iran~\cite{Goldman2012}, $hasPolMotive(iran, us, [2012, 2])$.
\hspace*{0pt}\hfill $\Box$
 \end{example}

\section{\emph{ABR}'s Background Knowledge}\label{sec:bg}
\emph{ABR} uses background knowledge comprising non-case-specific information and divided into \emph{general knowledge} and \emph{domain-specific knowledge}. 
Some of the background knowledge predicates used in this paper are shown
 in Table~\ref{table:bg}. 
 The use of the background knowledge alleviates the analysts' work and helps avoid human errors and bias.
 It comprises of pieces of information that
 are used as preconditions by the reasoning rules to answer the users' queries.
 \emph{ABR}'s background knowledge can be updated and enriched by the user. Note that reaching meaningful conclusions through the application of the rules to the background knowledge relies on the correctness of the background knowledge given. 

\begin{table}[htb] 
\centering
\scalebox{1}{
\begin{tabular}{|l|l|}
\hline
Predicate example & Explanation \\ \hline
\rowcolor[HTML]{FFFFC7} 
industry(infocomm) & Type predicate for industries\\ 
\rowcolor[HTML]{FFFFC7} 
polIndustry(military) & Political industries\\ 
\rowcolor[HTML]{FFFFC7} 
norIndustry(infocomm) & Non political industries\\ 
\rowcolor[HTML]{FFFFC7} 
country(united\_states) & Type predicate for countries \\
\rowcolor[HTML]{FFFFC7} 
cybersuperpower(united\_states) & List of cyber superpowers\\
\rowcolor[HTML]{FFFFC7} 
\begin{tabular}[c]{@{}l@{}}gci\_tier(afghanistan,initiating)\\ gci\_tier(poland, maturing)\\ gci\_tier(russian\_federation,leading)\end{tabular} & Global Cybersecurity Index (GCI)\\
\rowcolor[HTML]{FFFFC7} 
firstLanguage(english, united\_states) & First language used in the country \\
\rowcolor[HTML]{FFFFC7} 
goodRelation(united\_states, australia) & Good relations between countries \\
 \rowcolor[HTML]{FFFFC7} 
 poorRelation(united\_states, north\_korea) & Poor relations between countries \\
\rowcolor[HTML]{FFCE93} 
prominentGroup(fancyBear) & Prominent hacker groups \\
\rowcolor[HTML]{FFCE93} 
groupOrigin(fancyBear, russian\_federation) & Country of origin of a group \\
\rowcolor[HTML]{FFCE93} 
pastTargets(fancyBear, {[}france,...,poland{]}) & Past targets of a hacker group \\
\rowcolor[HTML]{FFCE93} 
malwareLinked(trojanMiniduke,cozyBear) & Past attribution of malware\\ 
\rowcolor[HTML]{FFCE93} 
malwareUsedInAttack(flame, flameattack) & \\
\rowcolor[HTML]{FFCE93} 
\begin{tabular}[l]{@{}l@{}}
ccServer(gowin7, flame) \\
\end{tabular} & 
C\&C servers of malware\\
\rowcolor[HTML]{FFCE93} 
\begin{tabular}[c]{@{}l@{}}domainRegisteredDetails(gowin7, \\ adolph\_dybevek, prinsen\_gate\_6)\end{tabular} & Dom. registration details of C\&C servers \\
 \hline
\end{tabular}}
\caption{Some of \emph{ABR}'s \emph{background knowledge}, divided into: general knowledge (yellow) and domain-specific knowledge (orange)\label{table:bg}}
\end{table}
\subsection{General Knowledge}
The general knowledge consists of information about countries characteristics,
 international relations between nations, and classification of the types of industry. 
 This information is used together with the given pieces of evidence, to perform the
 analysis.  
 Below we illustrate how these predicates are used by \emph{ABR}'s rules.

\emph{Language indicators} in malware can provide useful clues regarding the 
possible origin of attacks. 
We use two language artifacts: default system language settings, 
$sysLang$, and language used in code, $langInCode$. 
We present below two rules of \emph{ABR}, $t_5$ and $t_6$, that use the language 
evidence to derive the possible origin of the attack  $attackPOrig$, when the country's 
first language $firstLang$ matches the one found in the system/code.

\begin{displaymath}
\begin{array}{ll}
t_5: attackPOrig(X,Att) \leftarrow & sysLang(L, Att),\\ &
 firstLang(L,X).\\
t_6: attackPOrig(X,Att) \leftarrow & langInCode(L,Att),\\ & firstLang(L,X).
\end{array}
\end{displaymath}

The \emph{cyber capability of a nation} is another interesting information as it limits the type of attacks that an entity can carry out. 
We have used the Global Cybersecurity Index (GCI) Group~\cite{GCI2017} and  
the cyber capabilities of countries in cyberwar~\cite{WebBreene2016WhoSuperpowers} as sources for this information. There are three GCI groups: \emph{leading}, \emph{maturing} and \emph{initiating}, from where we classify the countries according to their capabilities. 
Furthermore, based on the cyber capabilities in 
cyberwar~\cite{WebBreene2016WhoSuperpowers} 
we identify some countries as cyber ``\emph{superpower}''.
We show below three of \emph{ABR}'s rules that use the countries' cyber capability. 

\begin{displaymath}
\begin{array}{l}
t_7: hasResources(X) \leftarrow  gci\_tier(X,leading).\\
t_8: hasResources(X) \leftarrow cybersuperpower(X).\\
t_9: hasNoResources(X) \leftarrow  gci\_tier(X,initiating).
\end{array}
\end{displaymath}
A country $hasResources$ if it is in the `leading' GCI group or is a cyber ``\emph{superpower}''.
Countries in the `initiating' GCI group are considered as $hasNoResources$.

\begin{example}\label{usbank2}
Let us continue with the $usBHack$ introduced in Example~\ref{usbank}.
In the background knowledge, we have that Iran is a cyber ``\emph{superpower}''\footnote{Iran is mentioned as a ``notable player" in~\cite{WebBreene2016WhoSuperpowers}. Thus, we identify it as a cyber ``superpower''.}. 
Thus, using rule $t_8$ ABR derives that Iran has the resources to carry out sophisticated attacks, $hasResources(iran)$.
Similarly the application of the operational rule $op_3$ derives that Iran has the capabilities\footnote{Note that we have a list of countries that have the resources to perform the attack, given their cyber capabilities. Rule $op_3$ is applied to all these countries. For the sake of simplicity, we only show the entities that are of interest to the discussed example.} to perform
the US bank hack, as shown below.
\begin{displaymath}
\begin{array}{ll}
op_3: hasCap(iran, usBHack) \leftarrow &
\  reqHighRes(usBHack), \\
& \ hasResources(iran).
\end{array}
\end{displaymath}
\hfill
 $\Box$
\end{example} 
Good \emph{international relations} between two countries can indicate that a state-sponsored attack is 
unlikely. 
We encoded this information in \emph{ABR} by creating a list of countries that have good
relations with each other ($goodRelation$) and a list of countries that may have poor relations
with each other ($poorRelation$) according to ~\cite{WebYouGov2017AmericasEnemies,WebBrilliantMaps2017WhoEnemies}.
This information can then be used to
 narrow down the countries that might have or not a motive to carry out an attack, as shown by the following rule.
\begin{displaymath}
\begin{array}{l}
op_5: \neg hasMotive(C,Att) \leftarrow  targetC(T,Att), \ country(T),\\ 
\qquad \qquad \qquad \qquad \qquad \quad \ \ \ country(C), goodRelation(C,T).
\end{array}
\end{displaymath}
Rule $op_5$ derives that country $C$ does not have any motive to perform $Att$,
as it has good relations with $T$ that is the country that $Att$ has as target, ($targetC(T, Att)$).

\subsection{Domain-Specific Knowledge}
Domain-specific knowledge consists of information about \emph{prominent 
groups} of attackers and \emph{past attacks}. 
These facts are primarily used in the \emph{strategic} and \emph{technical} layers.
 We encoded information on prominent APT groups
taken from~\cite{FireEye,Martin2016},
where for  each group we have their:
 name or ID;
country of origin;
countries/organisations targeted by the group in the past;
malware or pieces of malicious software (suspected or confirmed) linked to the group, as well as relations of this group with other entities (e.g., governments).
We assume these groups have the capabilities of conducting long term and significant 
attacks. 
Thus, we derive that an entity $X$ has the capabilities to perform
an attack, if $X$ is a prominent group of attackers, as shown by the rule below $op_6$. 
$$
op_6: hasCap(X,Att) \leftarrow prominentGroup(X).
$$

Another important part of the domain-specific knowledge is the
\emph{similarity}
with past attacks. For example, similarity to an APT-linked malware may indicate that the same APT group may be responsible. This is presented in rule $str_4$.
\begin{displaymath}
\begin{array}{ll}
str_4: \ isCulprit(X, A1) \leftarrow &malwareUsed(M1, A1),\\
& similar(M1, M2),\\ & malwareLinked(M2, X), \\
& notBlackMarket(M1), \\
&notBlackMarket(M2).\\
\end{array}
\end{displaymath}
In rule $str_4$, we derive that the attacker of $Att$ 
is most likely entity $X$,
because the malware used is similar to another malware linked to $X$,
and both malware codes were not found on the black market ($notBlackMarket$)\footnote{Currently, \emph{ABR} is not able to derive the evidence $notBlackMarket$. This evidence is either provided by the user or given in the background knowledge.}.
  We use the predicate $similar(M1, M2)$ to denote that two malwares are 
  similar to each other.
  In the rules below  we define that $M1$ and $M2$ are similar if
  they use a similar code obfuscation ($similarCodeObf$) mechanism, or they share code, or $M1$ is derived by modifying $M2$,
  or they have similar command and communication (C\&C) servers\footnote{For the sake of simplicity, in this paper we introduce only a subset of \emph{ABR}'s rules that  identify similarities between malwares or malicious software.}, where the similarity of C\&C servers of two different malwares can be derived by using other \emph{ABR}'s technical rules.
 \begin{displaymath}
\begin{array}{l}
  t_{10}: similar(M1, M2) \leftarrow similarCodeObf(M1, M2).\\
  t_{11}:  similar(M1, M2) \leftarrow  sharedCode(M1, M2).\\
  t_{12}:  similar(M1, M2) \leftarrow  modifiedFrom(M1, M2).\\
  t_{13}: similar(M1, M2) \leftarrow  similarCCServer(M1, M2).
\end{array}
\end{displaymath}

\section{Evaluation and Discussion}\label{discuss}

\emph{ABR} aims to be a flexible tool designed to be part of an iterative process, 
where the user can add other pieces of evidence, rules or preferences after evaluating the answers produced by the tool. \emph{ABR}'s input is given manually by the analyst, or could be collected, in part, automatically through an automatic extraction process by using digital forensics tools. 

\subsection{Evaluation}
We have tested \emph{ABR}'s performance and usability using examples of cyber-attacks published in the online literature.
During the evaluation, \emph{ABR} used the reasoning rules correctly to identify possible attackers. 
The explanations provided, in the textual and the graphical representations, helped to improve the usage of \emph{ABR} as they provided information that was used by the user for the next iterations. 
\emph{ABR} answered the queries requested 
(e.g., if a country had the motives or capabilities to perform the attack,
or if a particular group of attackers could be related to the attack following
the technical evidence of the used malware) as expected, given the pieces of evidence provided as input.
Note that \emph{ABR} assumes that the input evidence is correct, and providing inaccurate or incomplete information may lead to incorrect conclusions. 

For every tested example, we ran \emph{ABR} using a subset of the input evidence. 
Depending on the use-case and the provided evidence, \emph{ABR} was able to reach some 
conclusions by abducing (hypothesizing) some of the missing predicates. 
\emph{ABR} gave interesting results when asked to provide suggestions for missing evidence,
as it proposed useful missing evidence and also
new (not predicted) investigation paths. 
When a significant part of the evidence was provided to \emph{ABR}, its results coincided with those in the publicly available analyses or the entity attributed in the publicly available analyses was contained in \emph{ABR}'s list of possible culprits, which also contained other possibilities. 

Let us now briefly introduce some of the cyber-attacks used to evaluate \emph{ABR} and its conclusions. For the sake of space, we decided to show some well-known attacks where \emph{ABR} was tested, as they do not need a detailed introduction. 

\emph{ABR} analyzed evidence of the \emph{Stuxnet} attack~\cite{Zetter2014CountdownWeapon,mcafeeStuxnet} and derived two different entities as possible culprits: US and Israel. The \emph{Stuxnet} attack was first discovered in 2010 at the uranium enrichment plant in Iran. The code used was complex, using four zero-day vulnerabilities and mainly targeted Iran.
\emph{ABR} explained the conclusion based on the high resources required to perform such a sophisticated attack, and the political conflicts that existed in that period between Iran and US, and between Iran and Israel. In this case, the social evidence provided to \emph{ABR} mainly included the political conflicts between Iran and these two countries. However, ABR would have listed as possible source of the attacks any entity for which it could derive a motive and that had the resources to perform such sophisticated attacks. 

\emph{ABR} analyzed evidence of the \emph{Sony Pictures} attack~\cite{Roman2015FBIAttribution}. 
The Sony Pictures attack represents the 2014 attack when hackers infiltrated Sony's computers and stole data from Sony's servers. 
 A group called ``Guardians of Peace'' claimed credit for the attack, but several US government organisations claimed that the attack was state-sponsored by North Korea ~\cite{DoJ,GOP,Roman2015FBIAttribution}. 
\emph{ABR} attributed this attack to three possible culprits: the attackers group called ``Guardians of Peace''
and to two countries, Iran and North Korea. The attribution to Iran came as a consequence of low diplomatic relations between US and Iran. 
\emph{ABR}'s results were unexpected, with respect to the attribution given in~\cite{DoJ,GOP,Roman2015FBIAttribution}. In this case we see that \emph{ABR} can suggest new possible paths of investigation. 

\emph{ABR} analyzed evidence from the \emph{Conficker}~\cite{Conficker} attack and, using the evidence provided, was not able to reach a result.
Some of \emph{ABR}'s suggestions were to find entities that
operated/worked in Ukraine or that had interest in Ukraine, as
the first version of the attack was constructed to avoid machines with Ukrainian keyboards~\cite{confickerA}, thus to avoid a specific country (Ukraine). 
\emph{ABR} suggested also to find evidence about political or economical motivations for the attack, as the attack was sophisticated and it could either be a nation-state attack or performed by a cyber-criminal organization.

\begin{example}\label{ex3}
Let us now show \emph{ABR}'s final steps of the analysis and attribution of the $usBHack$. 
Following from Example~\ref{usbank} and \ref{usbank2}, \emph{ABR} derived the following predicates:
$hasCap$ $(iran, usBHack)$ and $hasPolMotive(iran, us, [2012, 2])$. 
\emph{ABR} can now apply the following operational rule:
\begin{displaymath}
\begin{array}{ll}
op_7: hasMotive(C, Att) \leftarrow & targetCountry(T, Att),\\ & attackPeriod(Att, Date1),\\ 
& hasPolMotive(C, T, Date2),\\ & specificTarget(Att)\\
& dateApplicable(Date1, Date2).
\end{array}
\end{displaymath}
Rule $op_7$ permits \emph{ABR} to derive that Iran has motives to perform the attack, as it has political motives.
Furthermore, these motives are applicable for the attack, as they occurred less than 1 year before the attack took place.
By applying rule $str_3$, \emph{ABR} derives that Iran is a possible culprit for this attack ($isCulprit(iran, usBHack)$). This result is in line with the attribution reported in~\cite{USBankByWhom},
while it does not match another attribution reported in~\cite{alQassam}, which attributed the attack to a group of hackers. 
\emph{ABR} provides its conclusion to the analyst together with its derivation tree with all the rules and evidence used. 

 \emph{ABR} provides further results of possible culprits, when new information is provided. For example, when new evidence is provided that a leader of the \emph{al-Qassam Cyber Fighters} hackers group has publicly claimed this attack~\cite{alQassam}, then \emph{ABR} derives that this group is also one of the possible attackers.

\hfill 
$\Box$
\end{example}

\subsection{Discussion}
Together with the answers to the queries, \emph{ABR} also provides 
the different ways in which the result was derived. 
Furthermore, every result comes with its \emph{explanation} that is composed of the rules and pieces of  evidence used.  
The explanations are in the form of text and graphical representation.
The given explanations make \emph{ABR}'s result and analysis process transparent 
to the user and provides her/him further information that can be used
for the analysis. ABR does not require the user to be familiar with the argumentation reasoning framework used as the user needs only to provide the evidence (in some cases the evidence is automatically extracted), and to launch the queries. 
The main goal of \emph{ABR} is to help the investigator during the analysis process and provide useful information. 

\emph{ABR}'s results include hypothesized but missing evidence and suggestions about other  investigations paths that could be followed by the analyst.
The missing evidence suggested can
be collected by the analyst in a second moment and given to \emph{ABR} as part 
of an iterative process. 
We decided to provide only the first list of 
results of the ``missing" pieces of evidence, together with the conclusions that could be derived from them,
 to keep \emph{ABR}'s running time and 
complexity polynomial.
Hence, \emph{ABR} does not 
provide an exhaustive list of all the possible hypotheses/missing evidence.
On the other hand, limiting the suggested evidence to be collected can be beneficial for the analyst, who can focus his/her attention on particular evidence, instead of spending time and resources on checking an exhaustive list. 

\emph{ABR} promotes best practice and helps to share lessons learned 
between analysts, as its reasoning rules can be constructed using the analysts'
reasoning process and be used by multiple investigators across different events. It also helps investigators cope with large numbers of rules and large knowledge bases. 
The attribution process is mainly human-based, and thus can be easily biased, e.g., by the resources invested~\cite{Felmlee}. 
In some cases, it may be difficult for the analyst to abandon a path of investigation when substantial resources have been devoted to it. 
\emph{ABR} permits to reduce the human bias through the rigorous application of rules and by suggesting new paths of investigation. 

\emph{ABR} relies on the reasoning rules with which it has been provided. Thus, \emph{ABR} can fail to deal with new evidence that has not been encountered before and which is not included in the reasoning rules. 
Furthermore, ABR relies on the rules being correct and complete. To illustrate its operation we have extracted 200 rules from public reports and analyses of cyber-attacks. However, the rules need to be validated with expert analysts and the rule base would need to be enriched with further rules for broader use. To facilitate the extraction and update of the reasoning rules and the background knowledge, we plan
 to investigate the automated extraction of rules and knowledge through the use of NLP techniques in conjunction with ontologies for cyber-attack investigations.

As \emph{ABR} uses in its reasoning information from past attacks and past attributions, it will derive the wrong conclusions if the information is incorrect. In particular, if a past attribution was incorrect, the error can be propagated to \emph{ABR}'s new results. 
For example, the Sony attack attribution~\cite{Roman2015FBIAttribution} was built on the (alleged) claim that 
North Korea was responsible for the assault on South Korean banks in 
2013~\cite{WebAltman2014SonyInterview}.
We can avoid this problem, by not using past attribution decisions as part of the knowledge, but it would make the attribution more difficult or cumbersome as it would require a larger amount of additional evidence. 
Furthermore, using results of past attributions is a common practice adopted by
forensic analysts during their analysis and attribution process,
as it permits to identify existing groups of attackers and to use their \emph{modus operandi}
as an important factor for the attribution.

\section{Conclusion and Future Work}
\label{concl}
In this work, we proposed a method and proof-of-concept argumentation-based reasoner (\emph{ABR})
that aims to help forensic investigators during the analysis and attribution process of cyber-attacks. Our aim was to demonstrate how such a tool can be constructed using the proposed argumentation framework.
 \emph{ABR} aims to help attribute cyber-attacks by leveraging both social and technical evidence. It provides explanations of the given results and hints of new investigation paths. 
The use of preference-based argumentation and abductive reasoning
permits \emph{ABR} to
work with conflicting pieces of evidence and
 to fill the knowledge gaps that derive from incomplete ones.
We introduced \emph{ABR}'s main components that are its reasoning rules
(that are based on past analyses and expert knowledge),
and its background knowledge.  
We improve \emph{ABR} usability by applying the Q-Model and categorizing the evidence and rules in three layers,
thus following a model familiar to the forensic analysts. 
Our reasoner emphasises the incremental and iterative nature of attribution,
by making the derivations of the solutions fully transparent to the user. 

In our future work, we plan to increase \emph{ABR}'s reasoning capabilities by 
adding new reasoning rules, and new background knowledge.
In this work, we mainly focused on constructing the \emph{ABR} reasoner, addressing its usability and showing its possible use. We leave a careful and possibly semi-automated population of the reasoning rules and background knowledge
for future work. In particular, we plan to use NLP techniques to automatically extract the reasoning rules and social evidence used by forensic analysts. 
Furthermore, we intend to enhance the expressive power of \emph{ABR}'s reasoner using ontologies. We also aim to address its integration with forensic tools and data mining techiniques.
 We plan to apply \emph{ABR} to other cyber-attacks, across a broad range of threats and to improve its usability using feedback from forensic analysts. 
 Another interesting future work is to include probabilities for our 
pieces of evidence and reasoning rules, in order to provide
probabilistic measures for the analysis and attribution results. 

\section*{Acknowledgments}
Erisa Karafili was supported by the European Union's H2020 research and innovation programme under the Marie Sk\l{}odowska-Curie grant agreement No.~746667.

\bibliographystyle{plain}
\bibliography{attribution}

\end{document}